\documentclass[sigconf,nonacm]{acmart}
\usepackage{multirow,tabularx,array,subcaption}
\AtBeginDocument{%
  }
\setcopyright{none}
\begin{document}

\title{LEDGE: An Efficient Simulator for Distributed LLM Serving on Heterogeneous Edge Devices}
\author{Zihao Wang}
\email{zihaow41@uci.edu}
\orcid{0009-0006-0897-1655}
\affiliation{%
  \institution{University of California, Irvine}
  \city{Irvine}
  \state{California}
  \country{USA}
}

\author{Ye Tian}
\email{yet002@ucsd.edu}
\orcid{0000-0002-8028-2532}
\affiliation{%
  \institution{University of California, San Diego}
  \city{San Diego}
  \state{California}
  \country{USA}
}

\author{Tajana Rosing}
\email{tajana@ucsd.edu}
\orcid{0000-0002-6954-997X}
\affiliation{%
  \institution{University of California, San Diego}
  \city{San Diego}
  \state{California}
  \country{USA}
}

\author{Nalini Venkatasubramanian}
\email{nalini@uci.edu}
\orcid{0000-0001-7011-2268}
\affiliation{%
  \institution{University of California, Irvine}
  \city{Irvine}
  \state{California}
  \country{USA}
}

\begin{abstract}
Evaluating distributed large language model (LLM) serving on heterogeneous edge devices requires exploring diverse model partitions, device placements, and network conditions, making repeated physical experiments costly. 
An efficient simulator for edge LLM serving must support diverse, fine-grained placement configurations, capture runtime behavior, and accurately predict execution times while keeping simulation overhead low. We present LEDGE, a lightweight simulator for fast, fine-grained evaluation of distributed LLM serving on heterogeneous edge devices without executing tensor operations.
LEDGE constructs event graphs supporting non-uniform tensor and pipeline parallelism, estimates execution costs using reusable performance models calibrated from end-to-end measurements, and simulates shared-link contention. 
We evaluate LEDGE on a heterogeneous testbed of 10 Jetson and Raspberry Pi devices, covering 17 model--precision combinations across 4 LLM families. On Qwen deployments, LEDGE achieves request-level mean absolute percentage errors below 3.6\% and a 135.3$\times$ aggregate speedup over physical inference. Performance model calibrated on one LLM family yield 7.44\% zero-shot end-to-end latency MAPE across three other families. Further simulations show how shared-link contention affects placement decisions and network topology constrains scaling. Across four edge environments, LEDGE completes 26,773 simulation runs in 83.8 minutes, enabling efficient exploration of deployment strategies under diverse resource conditions and serving objectives. The source code is publicly available.\footnote{\url{https://github.com/gdfwj/LEDGE}}
\end{abstract}

\maketitle
\section{Introduction}

Large language models (LLMs) are increasingly applied in mobile and pervasive computing systems, including personal health reasoning~\cite{tian2025dailyllm,tian2026lifeagentbench}, voice assistants~\cite{fang2025llama,defossez2024moshi}, and smart-home control~\cite{liu2025tasksense,king2024sasha}. Deploying LLMs at the edge can help preserve data privacy~\cite{zhang2026vcu,shen2025autoiot} and avoid the data transmission overhead of remote cloud inference. However, individual edge devices have limited computational and memory resources, constraining the models they can support and their inference performance. This motivates distributed LLM inference across multiple edge devices. Existing systems adopt different parallelization strategies to accommodate these constraints. For example, TPI-LLM~\cite{li2025tpi} employs tensor parallelism (TP) on compute-constrained devices, while PRIMA.cpp~\cite{li2026prima} uses pipeline parallelism (PP) for heterogeneous home clusters with limited memory and high-latency links.

Distributed LLM serving introduces a large deployment design space involving device selection, model partitioning, placement, and workload characteristics. Identifying efficient deployments requires considering not only the computation costs of different devices but also inter-device communication patterns and synchronization dependencies. Moreover, heterogeneous network conditions and shared links can cause interference between concurrent inference groups, making independently optimized deployments suboptimal when executed together. 
Evaluating these configurations through physical experiments is costly, particularly on resource-constrained devices with long inference times. 
Reproducing different device combinations and network conditions further compounds this cost. Therefore, before the actual deployment, efficient simulation is of great value for exploring the optimal candidate deployment.

\begin{figure}[t]
    \centering
    \includegraphics[width=0.47\textwidth]{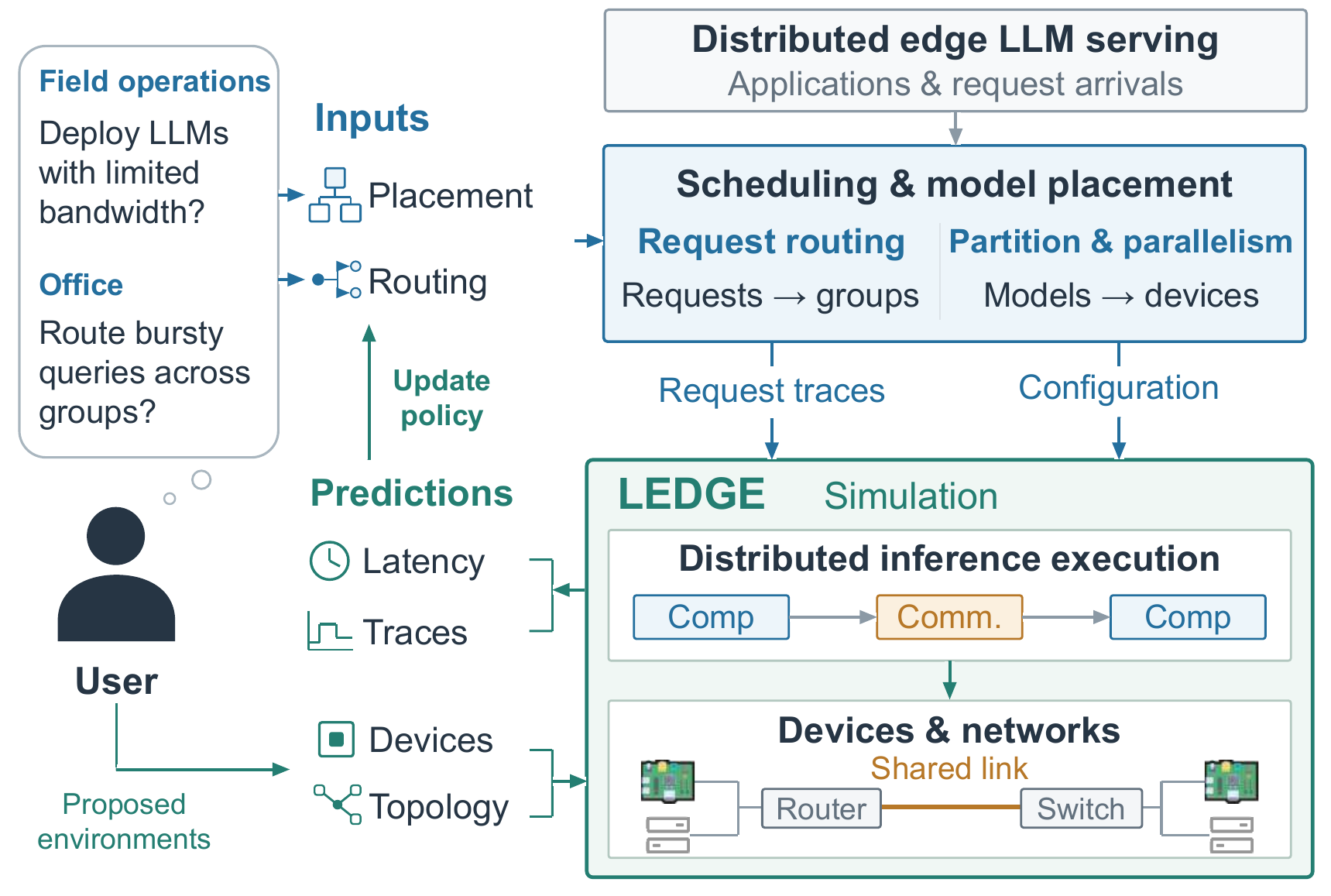}
    \caption{Evaluating placement and routing policies for distributed edge LLM serving with LEDGE.}
    \label{fig:intro}
    \vspace{-0.15in}
\end{figure}
Simulating distributed LLM serving at the edge presents three key challenges. First, model partitioning and device placement change the computation, communication, and synchronization dependencies, requiring accurate representation of different execution plans. Second, execution costs vary across devices, model precisions, and inference workloads, making accurate performance estimation difficult without extensive profiling. Third, concurrent inference groups may contend for shared network links, causing delays that propagate to dependent operations and affect subsequent resource contention. Addressing these challenges requires preserving timing-critical execution behavior while keeping simulation costs low.

Unfortunately, existing simulators do not fully address these challenges, leaving a gap in efficient and accurate evaluation of distributed LLM serving on heterogeneous edge devices. General-purpose edge simulators, such as EdgeCloudSim~\cite{sonmez2018edgecloudsim} and ncsim~\cite{krishnamachari2026ncsim}, model computation and network interactions but do not capture distributed LLM execution structures. Execution-based emulators, such as UNIFERENCE~\cite{eldenk2026uniference}, preserve application execution by performing actual model computations, limiting its simulation speed and the efficiency of repeated experiments.
LLM serving simulators, including Vidur~\cite{agrawal2024vidur} and LLMServingSim 2.0~\cite{cho2026llmservingsim}, avoid full model execution through profiling-based performance models. However, their predefined partitioning and accelerator-oriented performance models are not tailored to non-uniform deployments on heterogeneous edge devices. Its network simulation is also not tailored to capturing how shared-link contention across concurrent inference groups delays dependent operations and changes subsequent contention. These limitations motivate efficient and accurate simulation of distributed LLM serving on heterogeneous edge devices.

Our key idea is that computation is isolated across devices, while communication can contend for shared links. It allows us to directly estimate computation times and reserve detailed simulation for shared-link interactions, reducing simulation overhead while preserving accuracy.
Guided by this insight, we develop LEDGE, a lightweight simulator for distributed LLM serving on heterogeneous edge devices. LEDGE constructs compact event graphs that preserve computation--communication dependencies while supporting non-uniform TP/PP partitions and flexible device placement. Reusable performance models estimate event durations by separating workload demands from hardware-specific costs calibrated using end-to-end measurements, avoiding detailed operator-level profiling. A contention-aware discrete-event engine then advances concurrent inference groups on a shared timeline, capturing shared-link competition and the propagation of delays through dependent events. As illustrated in Fig.~\ref{fig:intro}, LEDGE takes user-defined deployment configurations and request traces as input and provides latency predictions and execution traces to support systematic deployment exploration.

We evaluate LEDGE on a heterogeneous testbed of ten Jetson and Raspberry Pi devices spanning four hardware types, with 17 model--precision combinations across Qwen, Llama, Mistral, and Yi. On Qwen deployments, LEDGE achieves request-level mean absolute percentage errors below 3.6\%, an event-duration mean absolute error of 1.70\,ms, and a 135.3$\times$ aggregate speedup over physical inference. Without additional calibration, the Qwen-calibrated parameters yield 7.44\% zero-shot end-to-end latency MAPE on Llama, Mistral, and Yi. Further simulations show how shared-link contention affects placement decisions and network topology influences scaling up to 64 devices. A case study across four application-inspired edge environments completes 26,773 simulation runs in 83.8 minutes, demonstrating efficient exploration of deployment strategies under diverse resource conditions and serving objectives. Our contributions are summarized as follows:

\begin{itemize}
    \item We design LEDGE, a lightweight simulator enabling rapid exploration of distributed LLM deployments on heterogeneous, resource-constrained edge devices without tensor execution.

    \item We develop a timing-centric simulation approach combining dependency-preserving event graphs, reusable performance models calibrated from end-to-end measurements, and contention-aware event simulation.

    \item We validate LEDGE on heterogeneous edge hardware, demonstrating prediction accuracy, simulation efficiency, cross-model generalization, and practical deployment exploration.
\end{itemize}

\section{Related Works}


\noindent\textbf{General-Purpose Edge Simulators.}
General-purpose edge simulators evaluate resource management and scheduling through abstract computation and communication models. EdgeCloudSim~\cite{sonmez2018edgecloudsim} and iFogSim~\cite{gupta2017ifogsim} support task offloading and placement, while iFogSim2~\cite{mahmud2022ifogsim2} and PureEdgeSim~\cite{mechalikh2021pureedgesim} extend support to mobility, service orchestration, and device heterogeneity. ncsim~\cite{krishnamachari2026ncsim} further incorporates DAG scheduling and wireless interference modeling. 
However, their abstractions for generic tasks lack support for processes in LLM inference, making them unsuitable for simulating distributed LLM serving.


\noindent\textbf{Execution-Based Edge Emulators.}
Execution-based edge emulators evaluate distributed applications under controlled device and network conditions. EmuFog~\cite{mayer2017emufog}, Fogify~\cite{symeonides2020fogify}, and MockFog 2.0~\cite{hasenburg2021mockfog} provide configurable fog environments for application deployment and evaluation. UNIFERENCE~\cite{eldenk2026uniference} combines threaded execution of distributed AI computations with discrete-event communication simulation. However, executing actual model computations incurs substantial compute and memory demands for LLM inference, making large-scale exploration of deployment configurations resource-intensive. 
In contrast, LEDGE simulates distributed LLM inference by estimating computation and communication times, which substantially reduces simulation time by avoiding actual model execution. 


\noindent\textbf{LLM Serving Simulators.}
LLM serving simulators estimate inference performance using profiling and performance models without full model execution. Vidur~\cite{agrawal2024vidur} explores scheduling, batching, and parallelism, while TokenSim~\cite{wu2025tokensim} and LLMServingSim~\cite{cho2024llmservingsim} support serving-policy and hardware exploration. LLMServingSim 2.0~\cite{cho2026llmservingsim} further supports heterogeneous accelerators and multi-model serving through operator-level profiles and execution models for serving stacks such as vLLM~\cite{kwon2023efficient}. However, these limited the flexibility of simulation, and their performance models are not tailored to the computation and communication patterns of heterogeneous edge systems. LEDGE is designed and evaluated for edge deployments, combining edge-specific computation models with contention modeling on shared links and fine-grained control over non-uniform model partitioning. It also calibrates performance parameters of new devices from end-to-end inference measurements without detailed operator-level profiling.

\section{LEDGE Design}
\label{sec:design}

\subsection{System Overview}
\label{sec:design-overview}

\begin{figure*}[t]
    \centering
    \includegraphics[width=1\textwidth]{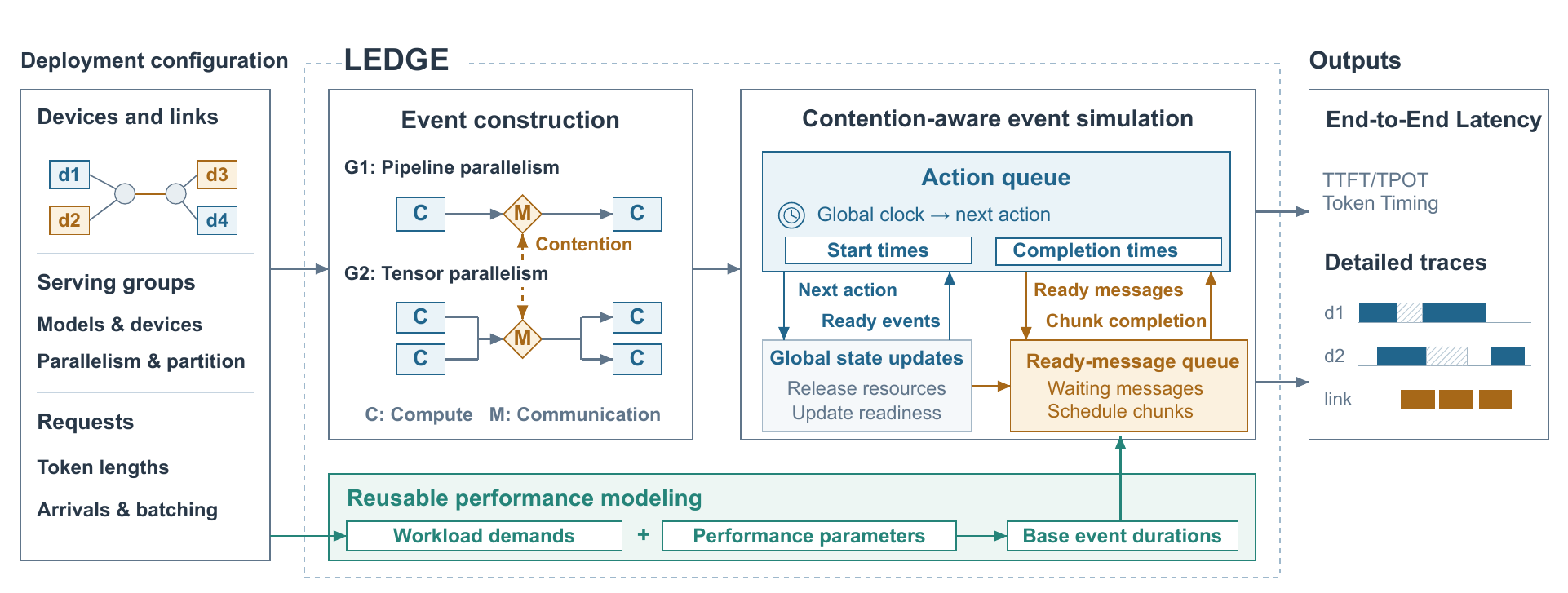}
  \caption{LEDGE system overview. The simulator predicts distributed LLM inference latency without executing tensor operations while preserving execution dependencies and shared-resource contention.}
    \label{fig:system-overview}
\end{figure*}


For a given workload and execution plan, predicting inference latency requires capturing execution costs and timing interactions rather than reproducing numerical results. Guided by this principle, LEDGE abstracts local computation into estimated execution costs while explicitly modeling dependencies and shared-link contention: dependencies govern when work becomes ready, while contention affects its progress and completion.
As shown in Fig~\ref{fig:system-overview}, LEDGE translates each model deployment and its requests into a graph of computation and communication events, preserving execution dependencies and resource mappings. Reusable performance models estimate base event durations from workload demands and calibrated device and network parameters. The simulator advances all serving groups on a shared timeline, coordinating event readiness and shared-link access so that contention-induced delays propagate to dependent events. LEDGE outputs request-level latency metrics and detailed execution traces.

\subsection{Fine-grained Event Construction}
\label{sec:design-event-construction}
Model partitioning determines both local workloads and the communication and synchronization needed to advance inference. LEDGE therefore aggregates local operators into computation events between communication and synchronization boundaries, preserving execution dependencies while keeping the event graph compact.
Each LLM instance and its assigned devices form a serving group. Within each group, PP connects stage computations through intermediate transfers, whereas TP connects concurrent local computations through collective rounds and synchronization dependencies. The event graph follows the execution plan across prefill microbatches and decoding iterations. Non-uniform PP layer assignments and TP partition ratios determine local workloads, while device placement and network topology determine the resources used by these events.

For each event, LEDGE records its predecessors, resource mappings, and workload demands derived from model structure, precision, partitioning, and request characteristics. A communication event represents a point-to-point transfer or a collective round and retains its constituent messages, including their sizes and routes, without expanding every message into a separate graph event. This representation exposes shared-link use for contention modeling. The resulting workloads are converted into base event durations by the performance models in Section~\ref{sec:design-performance}.

\subsection{Reusable Performance Modeling}
\label{sec:design-performance}

Model and deployment changes alter event workloads, while their execution costs depend on the target hardware and execution settings. LEDGE separates workload demands from calibrated hardware costs, allowing new configurations to reuse performance parameters by recomputing their workloads without repeating costly profiling. 

\noindent\textbf{Computation.}
LEDGE models arithmetic work, memory accesses, and graph-dispatch overhead. Arithmetic categories distinguish basic operations from matrix workloads, with the latter grouped by the number of tokens processed together to reflect kernel-efficiency differences across prefill, decoding, and microbatch sizes. For a local computation segment on device type $d$ with precision $q$, LEDGE estimates
\begin{equation}
    T_{\mathrm{comp}}
    = \sum_k x_{d,q,k} F_k + y_{d,q} D
    + z_{d,q} N_{\mathrm{graph}},
    \label{eq:computation-cost}
\end{equation}
where $F_k$ is the arithmetic work in category $k$, $D$ is the memory-access volume, and $N_{\mathrm{graph}}$ counts worker-graph dispatches. The calibrated coefficients $x_{d,q,k}$, $y_{d,q}$, and $z_{d,q}$ represent the effective costs of arithmetic work, memory access, and graph dispatch, respectively.

\noindent\textbf{Communication.}
For a message of size $V$ following path $p$, LEDGE estimates its communication cost in the absence of contention as
\begin{equation}
    T_{\mathrm{comm}}^{0}(V,p)
    =
    \sum_{\ell \in p} \alpha_{\ell}
    +
    \frac{V}{\min_{\ell \in p} B_{\ell}},
    \label{eq:communication-cost}
\end{equation}
where $\alpha_{\ell}$ and $B_{\ell}$ are the startup latency and effective bandwidth of link $\ell$. Startup latencies accumulate along the path, while its bottleneck bandwidth determines transfer time. 
During contention-aware simulation, path startup costs and shared-link transmission are accounted for separately to avoid double-counting transfer time.

\noindent\textbf{Calibration and reuse.}
LEDGE composes event costs according to execution dependencies and calibrates device and link parameters against measured TTFT and TPOT from isolated serving groups. Using both metrics across different workloads and partitions constrains prefill and decoding costs. New configurations are evaluated by recomputing workload demands and applying the corresponding device-type, precision, and link parameters, without requiring detailed operator-level profiling.

\subsection{Contention-Aware Event Simulation}
\label{sec:design-event-progression}

Shared-link contention can delay communication even when serving groups use different devices for computation. These delays postpone dependent work and alter subsequent link accesses, motivating LEDGE to evaluate execution dependencies and resource availability jointly.

\noindent\textbf{Dependency-driven progression.}
All serving groups advance on a shared timeline managed by a time-ordered action queue. The clock advances directly to the next scheduled state update, such as computation or chunk completion. Completing an event updates dependency states and releases successors whose prerequisites are satisfied, while independent work can proceed concurrently.

\noindent\textbf{Shared-link service.}
Inspired by packet queueing and interleaving on shared network links, LEDGE abstracts this behavior through chunk-level link service. To limit simulation overhead, LEDGE identifies potentially contending communication events during initialization and applies message-level scheduling only to these events. A ready-message queue arbitrates messages waiting for shared-link access. Each chunk occupies the required link resources until its transmission completes; completion releases those resources and returns an unfinished message to the queue, allowing competing messages to interleave. Messages using disjoint resources can progress concurrently. A communication event releases its successors only after its required messages complete, propagating communication delays to subsequent computation and transfers. For contending messages, chunk service accounts for the transfer term in Eq.~\ref{eq:communication-cost}, while path startup costs are charged separately.
\begin{figure}[t]
    \centering
    \includegraphics[width=\linewidth]{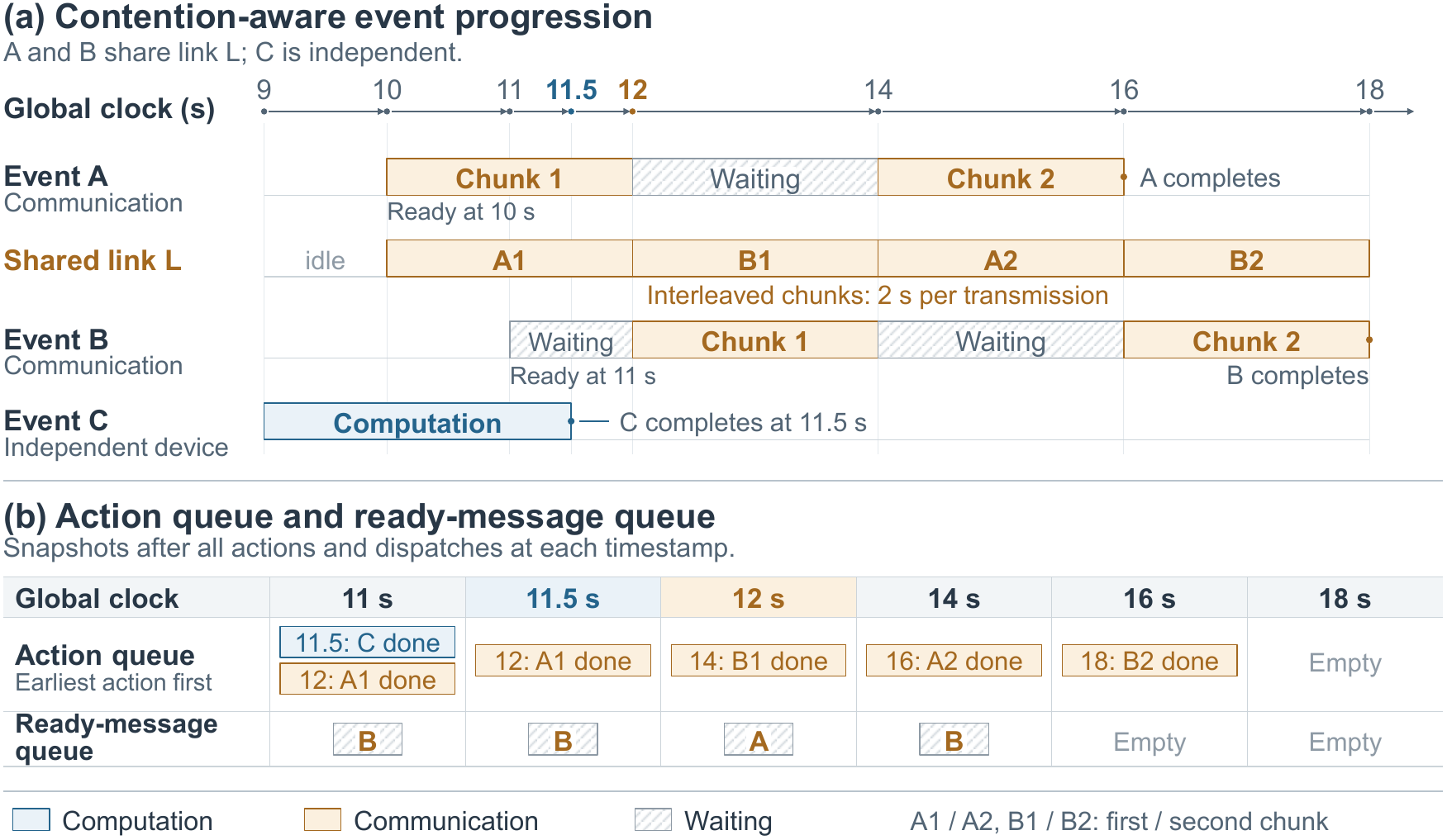}
    \caption{Event progression under shared-link contention.}
    \label{fig:contention}
\end{figure}
Figure~\ref{fig:contention} illustrates how two communication events alternate service on a shared link while an independent computation progresses on the same timeline.

\noindent\textbf{Tracing and replay.}
Fast and detailed modes use the same scheduling and contention model. Fast mode records request and token milestones; detailed mode additionally records computation, transmission, and waiting intervals for offline replay.

\section{Evaluation}

\subsection{Experimental Setup}

\noindent\textbf{Testbed and measurement.}
Our heterogeneous testbed comprises two NVIDIA Jetson Orin Nano boards, one Jetson Xavier, four Raspberry Pi 5 boards, and three Raspberry Pi 4 boards, with memory capacities of 2--32\,GB. Both Orin Nano boards and one Raspberry Pi 5 connect directly to the router via Gigabit Ethernet; the remaining devices connect through a switch with a shared uplink to the router. We obtain physical measurements using a customized \texttt{llama.cpp} backend supporting CPU and CUDA execution with MPI and NCCL communication, instrumenting graph-execution and communication interfaces to collect execution timings.

\noindent\textbf{Models and deployments.}
\label{model-config}
Physical validation covers 17 model--precision combinations across four families: Qwen3-4B, 8B, and 14B~\cite{yang2025qwen3}; Llama-3.2-3B and Llama-3.1-8B~\cite{grattafiori2024llama}; Mistral-7B~\cite{jiang2023mistral7b} and Mistral-Nemo-12B; and Yi-1.5-6B and 9B~\cite{young2024yi}. We evaluate both Q8\_0 and FP16 weights, except for Qwen3-14B, which is physically evaluated only with Q8\_0. Deployments span 2--10 devices with TP or PP and varying partitions. Additional simulation configurations are specified in the corresponding experiments.

\noindent\textbf{Workloads.}
Baseline fidelity measurements use synthetic requests with 32, 64, 128, or 256 input tokens and 16 or 32 output tokens. We construct inputs by repeating a fixed passage and truncating the tokenized sequence to the target length. EOS-based early stopping is disabled to enforce the specified output length, keeping workload sizes consistent across runs. Experiment-specific workloads are described in their respective sections.

\noindent\textbf{Evaluation protocol.}
We calibrate device and link parameters using measured time to first token (TTFT) and time per output token (TPOT) from isolated Qwen serving groups. Cross-model evaluation applies these parameters to Llama, Mistral, and Yi without additional calibration. 
At the request level, we evaluate predicted time to first token (TTFT), time per output token (TPOT), and end-to-end latency using mean absolute percentage error (MAPE) and P90 absolute percentage error. At the event level, we report duration mean absolute error (MAE) and P90 absolute error, together with weighted absolute percentage error (WAPE) for computation start times.

\subsection{Simulation Fidelity}
\label{evaluation:accuracy}
We first evaluate whether LEDGE accurately predicts distributed LLM inference timing without executing tensor operations. We test 63 Qwen configurations comprising 34 TP and 29 PP deployments, spanning 2, 3, 7, or 10 workers with varying partitions and CPU-only, GPU-only, or mixed CPU/GPU execution.

\noindent\textbf{Request-level accuracy.}
LEDGE achieves mean absolute percentage errors (MAPE) of 3.39\%, 3.29\%, and 3.59\% for time to first token (TTFT), time per output token (TPOT), and end-to-end (E2E) latency, respectively, with corresponding P90 absolute percentage errors of 7.66\%, 8.13\%, and 8.05\%. Notably, mixed CPU/GPU deployments achieve MAPE below 2.9\% across all three metrics, demonstrating accurate latency prediction under heterogeneous execution configurations.

\noindent\textbf{Event-level accuracy.}
Beyond end-to-end latency, we examine whether LEDGE accurately captures the event timing underlying distributed execution. Across 1.51 million matched computation and communication events, duration mean absolute error (MAE) is 1.70\,ms, with a P90 absolute error of 2.22\,ms. For 753,270 computation events, start-time weighted absolute percentage error (WAPE) is 4.80\%, while duration MAE and P90 absolute error are 1.18\,ms and 0.824\,ms, respectively. These results demonstrate that LEDGE captures both local execution costs and fine-grained timing across the evaluated distributed configurations.

\subsection{Simulation Efficiency}
\label{evaluation:efficiency}
We next evaluate whether replacing tensor execution with performance modeling substantially reduces the cost of exploring distributed LLM deployments.

\noindent\textbf{Evaluation speedup.}
We benchmark LEDGE on an Apple M5 across the 63 configurations in Section~\ref{evaluation:accuracy}, using the median runtime of seven API calls per configuration. Compared with physical inference, LEDGE achieves aggregate speedups of 136.9$\times$, 140.6$\times$, and 57.8$\times$ for Qwen3-4B, 8B, and 14B, respectively. The variation reflects differences in the evaluated workloads and deployments. Overall, LEDGE reduces 31.8 minutes of physical inference to 14.1 seconds of simulation, yielding a 135.3$\times$ aggregate speedup.

\begin{figure}[t]
    \centering
    \begin{subfigure}[t]{0.49\linewidth}
        \centering
        \includegraphics[width=\linewidth]{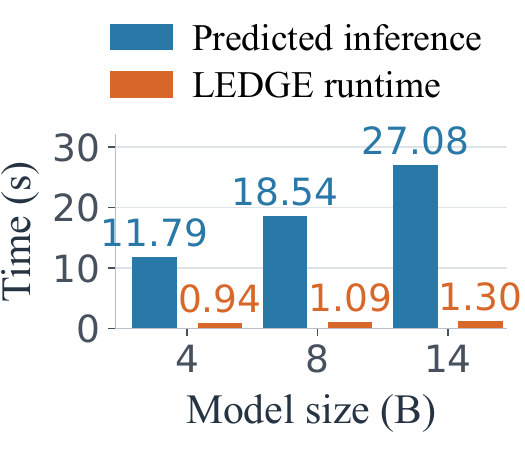}
        \caption{Varying model size}
        \label{fig:speedup-model-size}
    \end{subfigure}
    \hfill
    \begin{subfigure}[t]{0.49\linewidth}
        \centering
        \includegraphics[width=\linewidth]{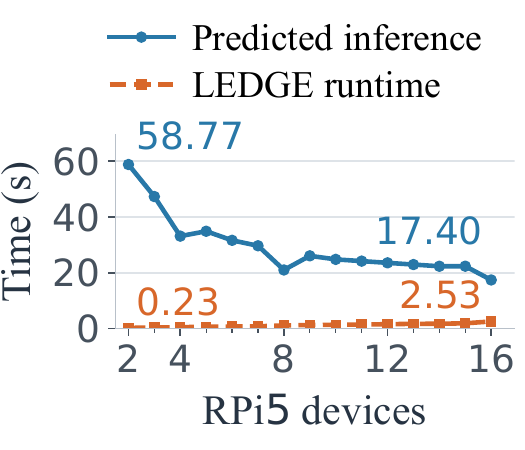}
        \caption{Varying deployment size}
        \label{fig:speedup-topology}
    \end{subfigure}
    \caption{Predicted inference time and LEDGE simulation runtime under varying model and deployment sizes.}
    \label{fig:simulator-speedup}
\end{figure}

\noindent\textbf{Varying Configurations.}
We further examine how simulation overhead changes with model and deployment size. Figure~\ref{fig:speedup-model-size} evaluates Qwen3-4B, 8B, and 14B with Q8\_0 weights. As model size increases from 4B to 14B, predicted inference time grows from 11.79 to 27.08 seconds, while simulation runtime increases only from 0.940 to 1.297 seconds. Figure~\ref{fig:speedup-topology} evaluates Qwen3-14B with FP16 weights across 2--16 Raspberry Pi 5 workers connected directly to a central router. To isolate simulation overhead, we disregard memory feasibility in this experiment. As the worker count increases, predicted inference time decreases from 58.77 to 17.40 seconds, while simulation runtime grows from 0.226 to 2.529 seconds due to the increased complexity of simulating communication among more workers. At 16 workers, simulation remains $6.9\times$ faster than the predicted inference time. Both experiments use requests with 128 input and 32 output tokens. These results show that LEDGE maintains low evaluation overhead as model and deployment sizes increase, enabling rapid exploration of configurations beyond the physical testbed.

\subsection{Cross-Model Generalization}
\label{evaluation:generalization}
The preceding experiments establish LEDGE's accuracy and efficiency on Qwen deployments. We further examine whether its calibrated performance parameters can be reused across model families without additional calibration. Specifically, we apply Qwen-calibrated parameters directly to newly measured Llama, Mistral, and Yi configurations with Q8\_0 and FP16 weights.
Across these configurations, zero-shot TTFT, TPOT, and E2E MAPE are 8.86\%, 4.53\%, and 7.44\%, respectively. E2E MAPE is 9.67\% for Llama, 4.25\% for Mistral, and 8.73\% for Yi. Despite increased errors relative to Qwen, E2E MAPE remains below 10\% across all three families. These results support the separation of workload demands from hardware-specific costs, enabling calibrated performance parameters to be reused across the evaluated dense LLM families without model-specific recalibration.

\subsection{Shared-Link Contention and Placement}
\label{evaluation:contention}
Placement decisions based on isolated inference costs may overlook interference among concurrent serving groups that share network resources. We use LEDGE to examine how shared-link contention affects placement choices even when inference groups use disjoint devices.
We simulate three two-worker TP groups on a heterogeneous testbed comprising two Jetson Orin devices ($O_1$, $O_2$), one Xavier ($X$), and three Raspberry Pi 5 devices ($P_1$, $P_2$, $P_3$). Devices $O_1$, $O_2$, and $P_3$ connect directly to a router, while $X$, $P_1$, and $P_2$ share a switch uplink. The groups serve Qwen3-4B Q8\_0, Qwen3-4B FP16, and Qwen3-8B Q8\_0 with 1, 32, and 8 sequences, respectively, each with 32 input and 32 output tokens.

We search memory-feasible placements using simulation and lower-bound pruning, comparing selections based on isolated group performance against those accounting for concurrent contention. Without cross-group contention, the selected placement is $(P_1,P_3)$, $(O_2,X)$, and $(O_1,P_2)$, with a mean end-to-end latency of 20.606\,s. However, its latency increases to 24.004\,s under concurrent execution. Accounting for shared-link contention instead selects $(P_2,P_3)$, $(O_1,O_2)$, and $(X,P_1)$, achieving 20.622\,s under the same concurrent workload, a 14.06\% reduction. This placement reduces the number of groups crossing the shared uplink from three to one.
These results show that shared-link contention can change preferred placements even without shared computing devices, highlighting the importance of modeling cross-group communication when evaluating distributed deployments.

\subsection{Network Topology and Scaling}
\label{evaluation:scaling}
\begin{figure}[t]
    \centering
    \includegraphics[width=0.65\linewidth]{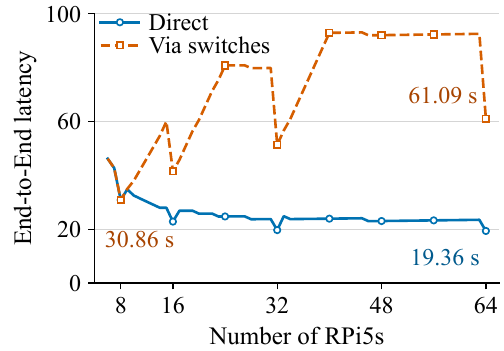}
    \vspace{-0.12in}
    \caption{Simulated end-to-end latency across worker counts under different network topologies.}
    \label{fig:scaling-fp16}
\end{figure}

Beyond contention among concurrent serving groups, communication within a distributed model can also limit the benefits of additional workers. We use LEDGE to examine how network topology affects the scaling of tensor-parallel inference.
We simulate Qwen3-14B with FP16 weights across 6--64 identical Raspberry Pi 5 workers, serving one request with 128 input and 64 output tokens. We compare workers connected directly to a central router against a switch-based topology, where each switch connects up to eight workers through a shared uplink.
Figure~\ref{fig:scaling-fp16} shows that direct connections exhibit diminishing returns: increasing the worker count from 32 to 64 reduces latency from 19.67 to 19.36\,s, an improvement of only 1.6\%. In contrast, the switch-based topology achieves its lowest latency of 30.86\,s at eight workers, but reaches 61.09\,s at 64 workers, with non-monotonic behavior due to additional communication rounds in MPI collectives. At 16 workers, collective communication accounts for 25.46\,s of the 41.55\,s end-to-end latency (61.3\%).
These results illustrate how communication overhead can offset the computational benefits of additional TP workers, making network topology a critical factor in scaling distributed edge LLM inference.

\section{Case Study}

\subsection{Scenarios and Search Space}
We use LEDGE to investigate how resource conditions, workloads, and serving objectives affect deployment choices for the same LLM. We consider four application-inspired environments: smart homes with interactive requests, offices with bursty queries, field operations with constrained connectivity, and infrastructure monitoring with periodic analysis tasks. All scenarios use Qwen3-8B with Q8\_0 weights on eight virtual workers across four sites in a star topology. Smart home and monitoring use calibrated Raspberry Pi 5 profiles, while office and field operation use Jetson Orin Nano profiles. We introduce device heterogeneity through performance scaling factors of 1--2. Table~\ref{tab:edge-environments} summarizes the network, workload, and optimization settings.
We explore a common space of 6,481 candidate deployments, varying worker selection, TP/PP configurations, and model partitioning. LEDGE evaluates feasible candidates under each environment's objective, including end-to-end latency, throughput, and deadline-constrained goodput. For monitoring, goodput counts requests completed within 30 minutes of arrival.

\subsection{Deployment Choices and Analysis}
\begin{figure}[t]
    \centering
    \includegraphics[width=0.95\linewidth]{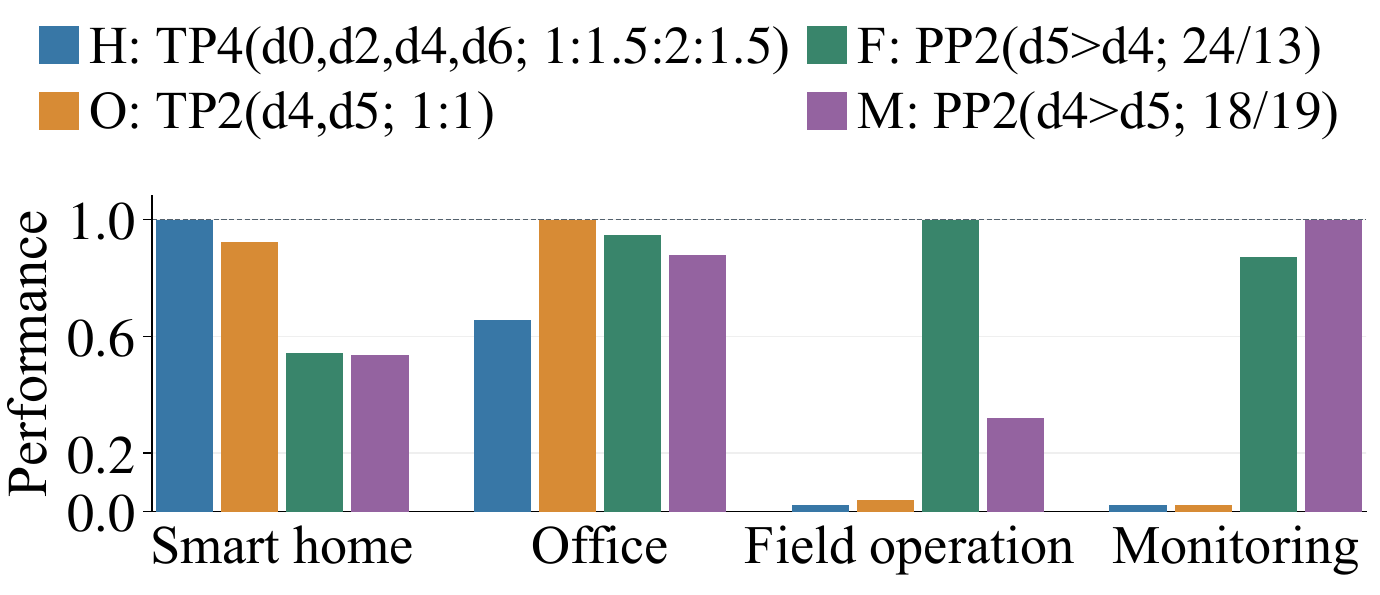}
    \caption{Performance across edge environments. Legend entries
    indicate placement, parallelism, and partitioning.}
    \label{fig:cross-scenario-rankings}
\end{figure}

The preferred deployments vary across environments. Smart home selects TP4, achieving 8.27\,s end-to-end latency, while office selects TP2, achieving 0.0346 requests/s throughput. Under constrained connectivity, field operation and monitoring select different PP2 configurations, achieving 6.40\,s end-to-end latency and 0.610 requests/min goodput, respectively. For monitoring, 47 of 48 requests meet the deadline.
Figure~\ref{fig:cross-scenario-rankings} compares these four deployments across all environments, with performance normalized to the best evaluated configuration in each scenario. Their relative performance changes substantially across environments, showing that a deployment effective under one set of conditions may perform poorly under another. These differences reflect varying computation and communication tradeoffs. Fast connectivity favors TP configurations in the evaluated scenarios, whereas constrained links increase the cost of frequent TP collectives. In field operation, the selected PP2 achieves a 26.4$\times$ lower end-to-end latency than the office-selected TP2. The two selected PP2 configurations also differ in stage order and layer allocation, consistent with different transfer requirements under their respective workloads.
The complete exploration requires 83.8 minutes for 26,773 simulation runs, representing 40.1 days of aggregate simulated inference time. This study demonstrates how LEDGE supports efficient deployment exploration across heterogeneous environments and helps identify the computation--communication tradeoffs underlying environment-specific choices.

\begin{table}[t]
    \centering
    \caption{Edge-environment configurations. Network values are shown as
    bandwidth (Mbps) / link startup latency (ms).}
    \label{tab:edge-environments}
    \footnotesize
    \setlength{\tabcolsep}{2pt}
    \begin{tabular}{@{}lccccc@{}}
        \toprule
        Scenario & Device & \shortstack{Network\\Mbps / ms} &
        \shortstack{Tokens\\In/Out} &
        \shortstack{Arrivals\\count (interval)} &
        Objective \\
        \midrule
        Smart home
        & RPi 5 & 1000/0.05 & 64/32
        & 1 & E2E $\downarrow$ \\

        Office
        & Orin Nano & 1000/0.05 & 256/512
        & \shortstack{16 (62.5 ms)}
        & Throughput $\uparrow$ \\

        \shortstack[l]{Field operation}
        & Orin Nano & 10/2 & 128/32
        & 1 & E2E $\downarrow$ \\

        \shortstack[l]{Monitoring}
        & RPi 5 & 10/2 & 1024/16
        & \shortstack{48 (60 s)}
        & Goodput $\uparrow$ \\
        \bottomrule
    \end{tabular}
\end{table}

\section{Conclusion}


We presented LEDGE, a lightweight simulator for distributed LLM serving on heterogeneous edge devices. LEDGE replaces tensor execution with reusable, calibrated performance models while preserving computation--communication dependencies and shared-link contention through event-based simulation. Evaluations on Jetson and Raspberry Pi devices demonstrate request-level latency errors below 3.6\% on Qwen deployments and a 135.3$\times$ aggregate speedup over physical inference, with calibrated parameters transferring across model families without additional calibration. Further simulations show how shared-link contention affects placement decisions, network topology constrains scaling, and resource conditions influence preferred partitioning strategies. By enabling systematic exploration of computation--communication tradeoffs across heterogeneous deployments, LEDGE provides a foundation for developing and evaluating future partitioning, placement, and resource-management strategies for distributed edge LLM serving.

\bibliographystyle{ACM-Reference-Format}
\bibliography{reference}

\appendix
\section{Performance Model Calibration}

\label{app:calibration}

LEDGE calibrates its performance model from measured TTFT and TPOT. Calibration covers the baseline costs of an isolated serving group and the contention parameters among concurrent groups. 

\subsection{End-to-End Cost Model}

The computation and communication cost equations in
Section~\ref{sec:design-performance} provide local execution
durations and communication durations per message. LEDGE combines these durations according to execution dependencies to predict TTFT and TPOT. We illustrate this composition using $n$-device TP and PP deployments, each processing one request from an idle state with a single prefill microbatch.

\noindent\textbf{Tensor parallelism.}
Let $\boldsymbol{\tau}=(\tau_1,\ldots,\tau_n)$ denote the
normalized tensor partition ratios, with $\tau_i>0$ and
$\sum_i\tau_i=1$. These ratios determine each device's local tensor slices through the backend's alignment rules. 
Assignments rotate across layers to distribute uneven slices; if the number of devices exceeds the number of attention heads, some devices will receive empty attention slices at every layer.
For synchronized TP execution, local computations proceed in parallel and join at collective boundaries. The duration of phase $\phi$, representing prefill or a decoding iteration, is
\begin{equation}
    T_{\mathrm{TP}}^\phi
    =
    H^\phi
    +
    \sum_j
    \left[
        \max_{1\leq i\leq n} C_{i,j}^\phi
        + M_j^\phi
    \right],
\end{equation}
where $C_{i,j}^\phi$ is the computation cost of segment $j$
on device $i$, calculated from its assigned workload and performance parameters. $M_j^\phi$ is the cost of the following collective, composed of its messages and rounds. The maximum captures synchronization with the slowest computation branch. $H^\phi$ accounts for controller processing and transfers before and after distributed execution.

\noindent\textbf{Pipeline parallelism.}
Consider devices ordered as $d_1,\ldots,d_n$, with device $d_i$ assigned $L_i$ of the model's $L$ Transformer layers. Its layer fraction is $\pi_i=L_i/L$, where $\sum_i L_i=L$ and $\sum_i\pi_i=1$.
For one prefill microbatch or decoding step, execution follows the stage order:
\begin{equation}
    T_{\mathrm{PP}}^\phi
    =
    H^\phi
    + \sum_{i=1}^{n} C_i^\phi
    + \sum_{i=1}^{n-1} M_{i\rightarrow i+1}^\phi.
\end{equation}
Here, $C_i^\phi$ includes the costs of the $L_i$ layers and additional operators assigned to stage $i$. $M_{i\rightarrow i+1}^\phi$ is the intermediate transfer cost along the path between consecutive stages. The layer fractions determine local workloads, while device assignments and stage order determine computation costs and communication paths.

For either deployment, prefill starts first and completes with the first generated token, while decoding continues until the final output token is generated. Time to first token (TTFT) measures prefill latency, while time per output token (TPOT) measures the average latency per subsequent token during decoding. A request producing $G>1$ tokens therefore has
\begin{equation}
    \mathrm{TTFT}=T^{\mathrm{prefill}},
    \qquad
    \mathrm{TPOT}
    =
    \frac{1}{G-1}
    \sum_{t=1}^{G-1}T_t^{\mathrm{decode}},
\end{equation}
where $T^{\mathrm{prefill}}$ is the prefill duration and $T_t^{\mathrm{decode}}$ is the elapsed time between the generation of the $t$-th and $(t+1)$-th output tokens.
Each decoding iteration uses workload quantities corresponding to its current context length. For multiple prefill microbatches and overlapping execution branches, LEDGE derives these durations from the execution timeline according to event dependencies.

\subsection{Parameter Fitting}

For each configuration $i$, consisting of a serving group and requests, we average request-level TTFT and TPOT over repetitions. For each metric $m \in \{\mathrm{TTFT}, \mathrm{TPOT}\}$, let $T_{i,m}$ denote the measured mean and $\widehat{T}_{i,m}(\boldsymbol{\theta}_q)$ the corresponding prediction. The relative residual is
\begin{equation}
    r_{i,m}(\boldsymbol{\theta}_q)
    =
    \frac{\widehat{T}_{i,m}(\boldsymbol{\theta}_q)-T_{i,m}}
         {T_{i,m}}.
\end{equation}
We fit each precision independently using the soft-$L_1$ objective
\begin{equation}
    \mathcal{L}_q
    =
    \sum_{i,m}\delta_q^2
    \left[
        \sqrt{1+
        \left(\frac{r_{i,m}}{\delta_q}\right)^2}
        -1
    \right],
\end{equation}
where $\delta_q$ controls the transition from quadratic to approximately linear growth. Each configuration contributes one residual per metric.

We jointly estimate computation and link parameters using bounded nonlinear least squares with an explicitly supplied Jacobian. Cost parameters are constrained to be nonnegative, and link parameters respect the link-capacity bound.
We test multiple initializations to explore different solutions. The final baseline selection additionally considers end-to-end latency error.

\subsection{Contention Calibration}
\label{sec:contention-calibration}

\noindent\textbf{Contention parameters.}
Inspired by packet interleaving, buffering, and contention-induced
delays in network communication, LEDGE uses a base service quantum
$Q$, sender-class weights $w_c$, an eager-transfer limit $E$, and
extra-latency parameters $\kappa_{\mathrm{prefill}}$ and
$\kappa_{\mathrm{decode}}$. To model interleaving, each message
receives up to $Qw_c$ bytes of service per turn and returns to the
queue if unfinished, with $w_c$ capturing differences in effective
sender behavior. To model buffering, up to $E$ bytes may be sent
before the receiver becomes ready. To account for additional
contention-induced delays, LEDGE adds
$\kappa_{\mathrm{phase}}W^{\mathrm{cross}}$ to message completion,
where $W^{\mathrm{cross}}$ is the waiting time caused by other
inference groups.

\noindent\textbf{Candidate search and calibration.}
We calibrate these parameters using measurements from inference runs with contention, holding the computation and base communication parameters fixed.
The main search fixes $Q=64$~KiB, sets Raspberry Pi sender weights to one, and preserves sender and receiver readiness constraints. We enumerate the following candidate values:
\begin{align*}
    E &\in \{0,Q\}, \\
    w_{\mathrm{Orin}},\,w_{\mathrm{Xavier}}
      &\in \{0.25,1,4\}, \\
    \kappa_{\mathrm{prefill}} &\in \{0,1,3,6\}, \\
    \kappa_{\mathrm{decode}} &\in \{0,0.5,1.5,3\}.
\end{align*}
For each candidate, we compare the TTFT and TPOT predicted by LEDGE
for all configurations with the corresponding measured means.
We select the candidate with the lowest mean squared relative error
across configuration groups.

We assess candidate selection through nine-fold grouped
cross-validation. Each configuration group contains the same model, deployment, and workload settings across different start offsets and repetitions. Each fold selects parameters using its training groups and evaluates them on the held-out groups. The deployed parameters are selected by repeating the search over all calibration groups.

\subsection{Calibrated Parameters and Interpretation}
\label{app:calibrated-parameters}

\begin{table}[t]
\centering
\small
\setlength{\tabcolsep}{3pt}
\caption{Representative calibrated parameters.}
\label{tab:calibrated-parameters}
\begin{tabular*}{0.9\columnwidth}{@{\extracolsep{\fill}}lrr@{}}
\toprule
Parameter (unit) & Q8\_0 & FP16 \\
\midrule
\multicolumn{3}{l}{\textit{Computation parameters}} \\
RPi4 $x$ (s/Gop)
    & 0.061443 & 0.028545 \\
RPi4 $y$ (s/GB)
    & 0.172067 & 0.247833 \\
RPi5 $x$, $t=1$ (s/Gop)
    & 0.044434 & 0.065788 \\
RPi5 $x_k$, $1<t\leq4$ (s/Gop)
    & 0.023557 & 0.022005 \\
RPi5 $x_k$, $8<t<64$ (s/Gop)
    & 0.009289 & 0.008215 \\
RPi5 $x_k$, $t\geq64$ (s/Gop)
    & 0.008665 & 0.008280 \\
RPi5 $y$ (s/GB)
    & 0.003590 & 0.028832 \\
Orin $x$ (s/Gop)
    & 0.00009319 & 0.00026007 \\
Orin $y$ (s/GB)
    & 0.014124 & 0.007381 \\
Xavier $x$ (s/Gop)
    & 0.00164833 & 0.00043685 \\
Xavier $y$ (s/GB)
    & 0.007350 & 0.002579 \\
\midrule
\multicolumn{3}{l}{\textit{Communication parameters}} \\
$\alpha_\ell$ range ($\mu$s)
    & 0.62--181.68 & 1.34--346.20 \\
$\alpha_{\mathrm{uplink}}$ ($\mu$s)
    & 2.85 & 8.48 \\
$b_\ell$ range (s/GB)
    & 8.00--10.92 & 8.00--11.81 \\
$b_{\mathrm{uplink}}$ (s/GB)
    & 8.00 & 8.00 \\
\midrule
\multicolumn{3}{l}{\textit{Contention settings}} \\
$Q$ (KiB, fixed)
    & \multicolumn{2}{c}{64} \\
$E$ (KiB)
    & \multicolumn{2}{c}{64} \\
$w_{\mathrm{RPi}}$ (fixed)
    & \multicolumn{2}{c}{1} \\
$w_{\mathrm{Orin}}$
    & \multicolumn{2}{c}{4} \\
$w_{\mathrm{Xavier}}$
    & \multicolumn{2}{c}{0.25} \\
$\kappa_{\mathrm{prefill}}$
    & \multicolumn{2}{c}{1} \\
$\kappa_{\mathrm{decode}}$
    & \multicolumn{2}{c}{0} \\
Rank readiness
    & \multicolumn{2}{c}{Enabled} \\
\bottomrule
\end{tabular*}
\end{table}

Table~\ref{tab:calibrated-parameters} summarizes some representative
calibrated parameters. Gop denotes $10^9$ operations.

\noindent\textbf{Computation parameters.}
Base arithmetic costs for Orin and Xavier are substantially lower than those for Raspberry Pi devices, while both arithmetic costs and memory-access costs for RPi5 are lower than those for RPi4, capturing the heterogeneity across devices.
For RPi5, the arithmetic parameter decreases substantially from
$0.044434$ to $0.008665$~s/Gop under Q8\_0 and from $0.065788$
to $0.008280$~s/Gop under FP16 when comparing $t=1$ with
$t\geq64$. This change reflects different computation paths and higher effective arithmetic efficiency at larger $t$. Consistent with this observation, RPi4 has the largest memory-access parameter under both precisions.

\noindent\textbf{Communication parameters.}
The fitted startup delays are below $0.35$~ms, and the uplink startup delays are particularly small: $2.85$~$\mu$s for Q8\_0 and $8.48$~$\mu$s for FP16. With an uplink service cost of $8$~s/GB, consistent with the negotiated bandwidth of $1$~Gbps, a 64~KiB message has a byte-transfer cost of approximately $0.524$~ms.
Startup therefore contributes approximately $0.54\%$ and $1.59\%$ of the message's uplink cost, respectively. Transfer cost dominates on this link for messages of this
size or larger. For smaller messages, startup contributes a larger fraction of
communication time and accumulates across hops and collective rounds.

\noindent\textbf{Contention parameters.}
The selected sender weights yield service quanta of up to
256~KiB for Orin, 64~KiB for Raspberry Pi, and 16~KiB for Xavier,
capturing different effective service granularities among senders.
The extra-latency parameters $\kappa_{\mathrm{prefill}}=1$ and $\kappa_{\mathrm{decode}}=0$ retain an additional waiting-dependent completion delay during prefill, consistent with the observation that prefill consumes a larger share of network resources~\cite{jin2025distributed}. 
During decoding, contention delays arise from the explicitly
modeled shared-link queueing and readiness constraints.

\section{Additional Experimental Results}

\begin{figure}[t]
  \centering
  \includegraphics[width=\linewidth]{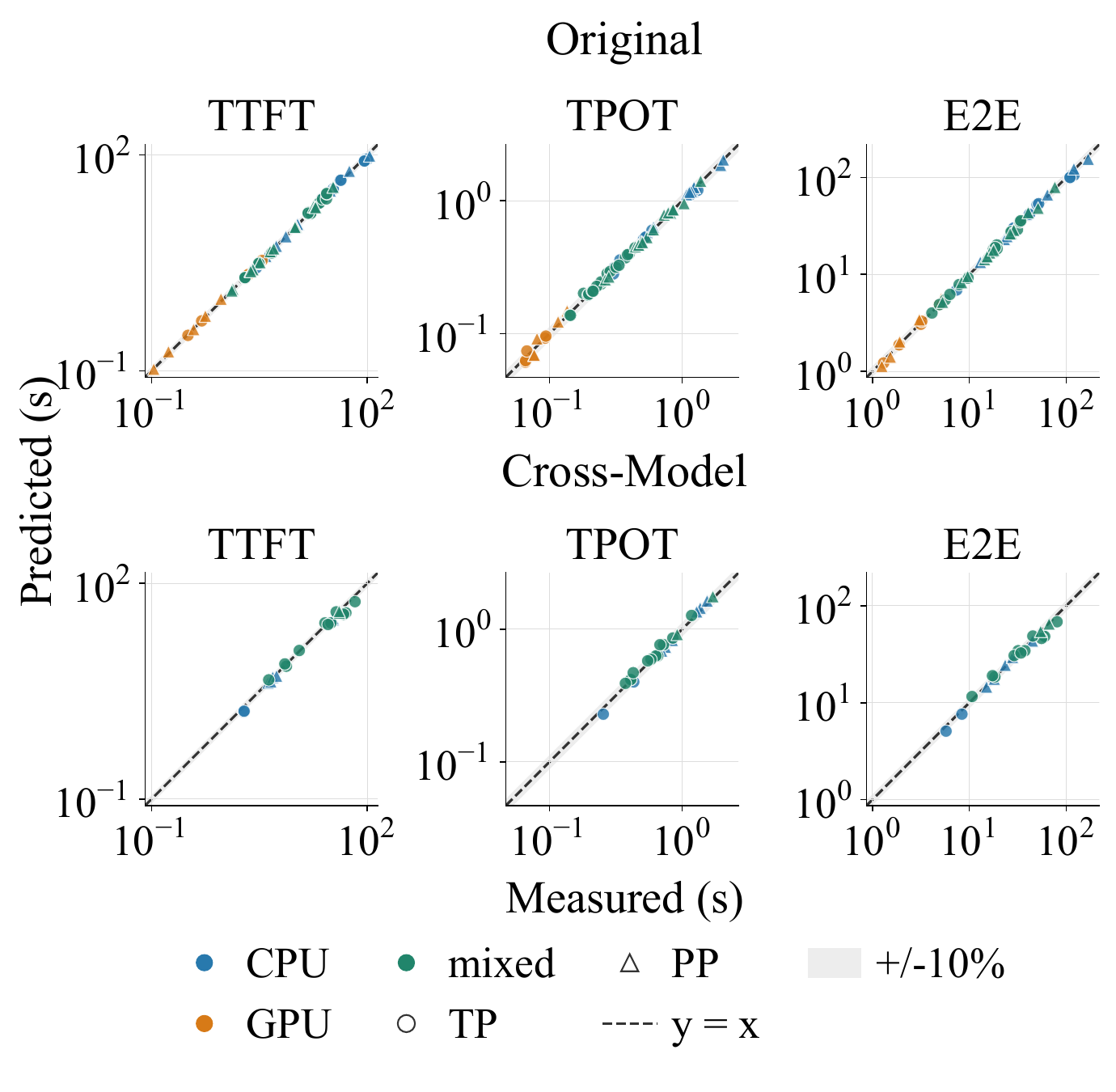}
  \caption{Measured and predicted request latency for Qwen calibration configurations (top) and cross-model configurations (bottom). The dashed line marks $y=x$ and the shaded band spans $0.9x$ to $1.1x$.}
  \label{fig:append-fidelity-scatter}
\end{figure}

We provide further analysis of the previous experimental results. Figure~\ref{fig:append-fidelity-scatter} compares measured and predicted latency for the Qwen configurations used for calibration in Section~4.2 and the cross-model configurations evaluated with the same parameters in Section~4.4. For E2E latency, 61 of 63 Qwen configurations and 17 of 22 cross-model configurations fall within 10\% of measured latency. Table~\ref{tab:append-main-fidelity-groups} shows where the cross-model predictions differ most from the measurements. TTFT has a MAPE of 8.86\% and a P90 APE of 18.71\%, compared with 4.53\% and 9.79\% for TPOT.

For Qwen, Table~\ref{tab:append-main-fidelity-groups} shows similar E2E MAPE for TP and PP (3.81\% and 3.32\%). Mixed CPU/GPU deployments have the lowest E2E MAPE among the three worker groups (2.54\%). In the cross-model experiment, the gap between TP and PP is larger for TTFT (11.68\% versus 3.94\%) than for TPOT (5.19\% versus 3.35\%). 

For cross-model generalization, the family and precision rows show a similar difference between prefill and decode. TTFT MAPE is higher than TPOT MAPE for Llama and Yi, while Mistral has lower TTFT MAPE (4.24\%) than TPOT MAPE (5.79\%). Across the two precisions, FP16 has higher TTFT MAPE (10.95\% versus 6.77\%), while TPOT MAPE is similar (4.28\% versus 4.77\%). 

\begin{figure}[t]
  \centering
  \includegraphics[width=\linewidth]{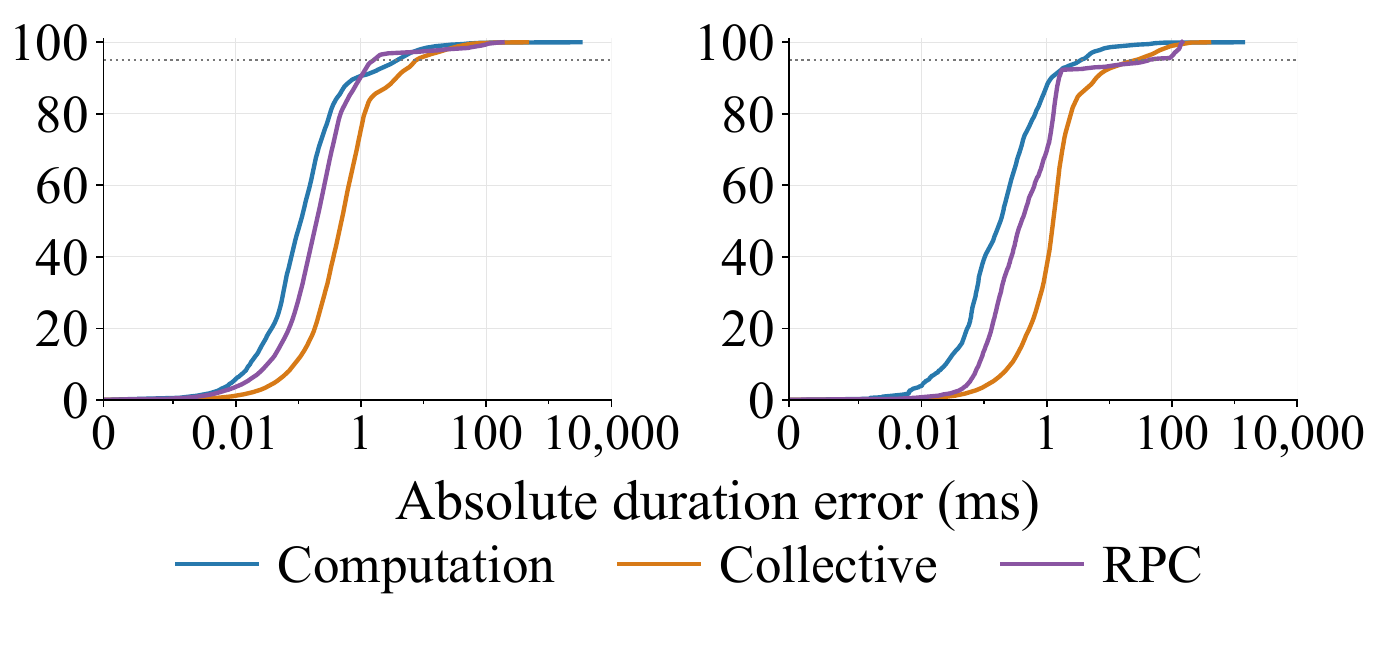}
  \caption{Measured and predicted request latency for Qwen calibration configurations (top) and cross-model configurations (bottom). The dashed line marks $y=x$ and the shaded band spans $0.9x$ to $1.1x$.}
  \label{fig:append-fidelity-event-diagnostics}
\end{figure}

Figure~\ref{fig:append-fidelity-event-diagnostics} shows how absolute duration errors vary across individual events. Computation duration MAE is similar for Qwen and the cross-model configurations (1.18 and 1.08\,ms), while communication duration MAE is higher for the cross-model configurations (5.56 versus 2.22\,ms). In the Qwen panel, 95\% of RPC payload events have an absolute duration error of at most 1.59\,ms, demonstrating the fidelity of LEDGE.

\begin{table}[t]
\centering
\caption{Fidelity in the original and cross-model experiments.}
\label{tab:append-main-fidelity-groups}
\footnotesize
\setlength{\tabcolsep}{1.6pt}
\resizebox{0.85\columnwidth}{!}{%
\begin{tabular}{@{}llr rr rr rr@{}}
\toprule
\multirow[c]{2}{*}{Grouping} & \multirow[c]{2}{*}{Subset}
 & \multirow[c]{2}{*}{$N$} & \multicolumn{2}{c}{TTFT}
 & \multicolumn{2}{c}{TPOT} & \multicolumn{2}{c}{E2E} \\
\cmidrule(lr){4-5}\cmidrule(lr){6-7}\cmidrule(lr){8-9}
 & & & MAPE & P90 & MAPE & P90 & MAPE & P90 \\
\midrule
\multicolumn{9}{@{}l}{\textbf{Original (Qwen)}} \\
\specialrule{0.3pt}{2pt}{2pt}
N/A & All & 63 & 3.39 & 7.66 & 3.29 & 8.13 & 3.59 & 8.05 \\
\specialrule{0.3pt}{2pt}{2pt}
\multirow[c]{2}{*}{Parallelism} & TP & 34 & 4.06 & 9.07 & 2.94 & 7.49 & 3.81 & 7.46 \\
 & PP & 29 & 2.62 & 7.66 & 3.70 & 10.07 & 3.32 & 8.99 \\
\specialrule{0.3pt}{2pt}{2pt}
\multirow[c]{3}{*}{Workers} & CPU-only & 19 & 4.61 & 9.63 & 3.08 & 7.27 & 4.40 & 8.62 \\
 & GPU-only & 10 & 3.00 & 7.66 & 7.33 & 10.30 & 5.59 & 9.55 \\
 & Mixed & 34 & 2.83 & 6.90 & 2.22 & 4.49 & 2.54 & 5.62 \\
\specialrule{0.3pt}{2pt}{2pt}
\multirow[c]{2}{*}{Precision} & Q8\_0 & 49 & 4.09 & 9.07 & 4.17 & 9.53 & 4.40 & 8.62 \\
 & FP16 & 14 & 0.96 & 3.33 & 0.22 & 0.33 & 0.72 & 2.73 \\
\midrule
\multicolumn{9}{@{}l}{\textbf{Cross-model}} \\
\specialrule{0.3pt}{2pt}{2pt}
N/A & All & 22 & 8.86 & 18.71 & 4.53 & 9.79 & 7.44 & 16.18 \\
\specialrule{0.3pt}{2pt}{2pt}
\multirow[c]{2}{*}{Parallelism} & TP & 14 & 11.68 & 20.70 & 5.19 & 10.10 & 10.00 & 17.19 \\
 & PP & 8 & 3.94 & 7.29 & 3.35 & 5.84 & 2.97 & 5.21 \\
\specialrule{0.3pt}{2pt}{2pt}
\multirow[c]{3}{*}{Workers} & CPU-only & 8 & 6.83 & 15.16 & 5.02 & 10.10 & 5.22 & 11.36 \\
 & GPU-only & 0 & -- & -- & -- & -- & -- & -- \\
 & Mixed & 14 & 10.03 & 20.70 & 4.24 & 9.79 & 8.71 & 17.19 \\
\specialrule{0.3pt}{2pt}{2pt}
\multirow[c]{3}{*}{Family} & Llama & 8 & 12.52 & 22.88 & 4.20 & 10.10 & 9.67 & 19.38 \\
 & Mistral & 8 & 4.24 & 6.48 & 5.79 & 12.09 & 4.25 & 9.48 \\
 & Yi & 6 & 10.16 & 20.70 & 3.27 & 5.84 & 8.73 & 17.19 \\
\specialrule{0.3pt}{2pt}{2pt}
\multirow[c]{2}{*}{Precision} & Q8\_0 & 11 & 6.77 & 10.05 & 4.77 & 10.10 & 6.07 & 9.48 \\
 & FP16 & 11 & 10.95 & 20.70 & 4.28 & 7.22 & 8.82 & 17.19 \\
\bottomrule
\end{tabular}%
}
\end{table}

\section{Additional Experiments on Fidelity}
\label{app:additional-fidelity}

\noindent\textbf{Setup.}
To demonstrate LEDGE's fidelity across a wide range of real-world settings, we further test LEDGE on additional deployments and workloads.
For the deployment experiments, each request contains one sequence with 128 input tokens and 32 output tokens. We test TP and PP deployments with 2--6 workers, varying the partitions, device assignments, and, for PP, stage order.
For the workload experiments, we use eight four-worker TP deployments. We test batches of 4, 6, or 8 sequences per request, longer inputs of 512 or 1,024 tokens, and longer outputs of 128 or 256 tokens in separate experiments. We also test prefill microbatch limits of 48, 64, or 96 tokens with eight 32-token input sequences per request.
For each configuration, we average three timed runs after three warmups. Following Section~4.2, we report MAPE and P90 absolute percentage error (APE) for TTFT, TPOT, and E2E latency.

\noindent\textbf{Results.}
According to Table~\ref{tab:additional-fidelity}, LEDGE achieves MAPE of 7.89\% for TTFT, 7.10\% for TPOT, and 5.36\% for E2E latency. E2E predictions fall within 10\% of measured latency in 194 of 224 configurations. The two settings yield similar E2E MAPE (5.26\% and 5.57\%). Across new deployments, TTFT has higher error than TPOT (8.32\% versus 6.32\%). Across new workloads, TPOT has higher error (8.66\% versus 7.04\%).

\begin{figure}[t]
\centering
\includegraphics[width=\columnwidth]{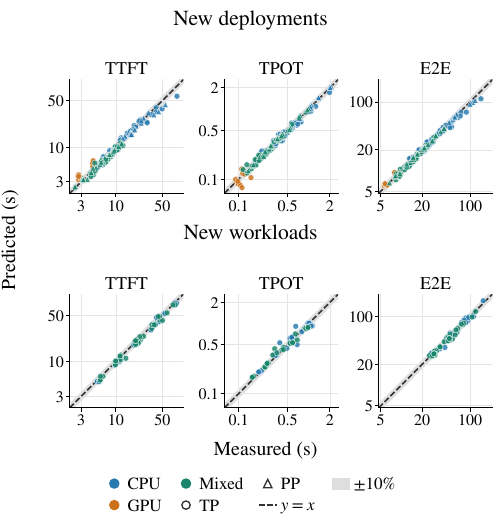}
\caption{Measured and predicted request latency for new deployments (top) and new workload configurations (bottom). The dashed line marks $y=x$ and the shaded band spans $0.9x$ to $1.1x$.}
\label{fig:additional-fidelity}
\end{figure}

\begin{table}[t]
\centering
\caption{Request-level fidelity on the new configurations. $N$ counts configurations. MAPE and P90 APE are percentages.}
\label{tab:additional-fidelity}
\footnotesize
\setlength{\tabcolsep}{1.6pt}
\resizebox{0.85\columnwidth}{!}{%
\begin{tabular}{@{}llr rr rr rr@{}}
\toprule
\multirow[c]{2}{*}{Grouping} & \multirow[c]{2}{*}{Subset}
 & \multirow[c]{2}{*}{$N$} & \multicolumn{2}{c}{TTFT}
 & \multicolumn{2}{c}{TPOT} & \multicolumn{2}{c}{E2E} \\
\cmidrule(lr){4-5}\cmidrule(lr){6-7}\cmidrule(lr){8-9}
 & & & MAPE & P90 & MAPE & P90 & MAPE & P90 \\
\midrule
N/A & All & 224 & 7.89 & 14.83 & 7.10 & 13.93 & 5.36 & 11.02 \\
\specialrule{0.3pt}{2pt}{2pt}
\multirow[c]{2}{*}{Evaluation} & Devices & 149 & 8.32 & 17.49 & 6.32 & 13.15 & 5.26 & 11.21 \\
 & Workloads & 75 & 7.04 & 10.88 & 8.66 & 15.61 & 5.57 & 11.02 \\
\specialrule{0.3pt}{2pt}{2pt}
\multirow[c]{2}{*}{Parallelism} & TP & 162 & 8.79 & 15.78 & 8.60 & 16.42 & 6.33 & 11.45 \\
 & PP & 62 & 5.55 & 8.81 & 3.18 & 5.66 & 2.84 & 4.70 \\
\specialrule{0.3pt}{2pt}{2pt}
\multirow[c]{3}{*}{Workers} & CPU-only & 122 & 7.28 & 12.01 & 7.33 & 14.28 & 5.88 & 11.21 \\
 & GPU-only & 10 & 23.20 & 36.69 & 15.13 & 29.94 & 7.53 & 13.42 \\
 & Mixed & 92 & 7.04 & 13.62 & 5.93 & 10.31 & 4.44 & 9.32 \\
\specialrule{0.3pt}{2pt}{2pt}
\multirow[c]{2}{*}{Model size} & 4B & 120 & 7.30 & 13.40 & 7.49 & 14.28 & 4.90 & 9.87 \\
 & 8B & 104 & 8.58 & 16.01 & 6.65 & 13.15 & 5.91 & 11.12 \\
\specialrule{0.3pt}{2pt}{2pt}
\multirow[c]{2}{*}{Precision} & Q8\_0 & 120 & 7.54 & 13.97 & 6.48 & 13.47 & 4.99 & 10.30 \\
 & FP16 & 104 & 8.30 & 18.99 & 7.82 & 14.66 & 5.80 & 11.34 \\
\bottomrule
\end{tabular}%
}
\end{table}

\noindent\textbf{Analysis.}
Figure~\ref{fig:additional-fidelity} shows the differences between measured and predicted latency. For PP, 61 of 62 configurations fall within the 10\% E2E band, compared with 133 of 162 for TP, demonstrating that TP's more complex communication patterns make prediction and generalization more challenging. The GPU-only points show that LEDGE overpredicts TTFT in all ten GPU-only configurations and underpredicts TPOT in seven, which may result from fewer calibration data. However, the GPU-only E2E MAPE is 7.53\%, although TTFT and TPOT MAPE reach 23.20\% and 15.13\%, respectively. Mixed CPU/GPU execution has lower errors across all three metrics, with E2E MAPE of 4.44\% across 92 configurations. 

\section{Experiments on Contention Modeling}
\label{app:contention-validation}

\noindent\textbf{Setup.}
We test whether LEDGE predicts how much a serving group slows down when another group uses the same link. We set up a testbed with 3 serving groups. Groups A and B run on separate workers but both send model data over the link between the router and the switch. Group C sends model data only between workers on the switch. In the original experiment, we run each group alone and then run two or three groups together. We test requests with 128 input and 32 output tokens, 512 input and 32 output tokens, or 128 input and 128 output tokens. The extended experiment varies weight precision (Q8\_0 or FP16), batch size (one or two sequences), prefill microbatch size, and group start times. We use the parameters from Appendix~A and compare LEDGE with a version that disables cross-group contention. This version keeps the same communication paths and costs but gives each group its own link queues. For each group, we measure TTFT and TPOT from its own request start and average three timed runs after three warmups. We define slowdown as concurrent latency divided by the same group's isolated latency, computing this ratio separately for measurements and predictions. 

\noindent\textbf{Results.}
Table~\ref{tab:contention-validation} shows that enabling cross-group contention lowers both TTFT and TPOT latency MAPE in both experiments: MAPE falls from 2.59\% to 1.62\% in the original experiment and from 2.69\% to 1.80\% in the extended experiment, showing the improvement from contention modeling. 

\begin{table}[t]
\centering
\caption{Prediction error with cross-group contention enabled (On) or disabled (Off).}
\label{tab:contention-validation}
\footnotesize
\setlength{\tabcolsep}{3pt}
\resizebox{0.85\columnwidth}{!}{%
\begin{tabular}{@{}llr rr rr@{}}
\toprule
\multirow[c]{2}{*}{Experiment} & \multirow[c]{2}{*}{Quantity} & \multirow[c]{2}{*}{$N$}
 & \multicolumn{2}{c}{TTFT} & \multicolumn{2}{c}{TPOT} \\
\cmidrule(lr){4-5}\cmidrule(lr){6-7}
 & & & Off & On & Off & On \\
\midrule
\multirow[c]{2}{*}{Original} & Latency & 31 & 7.06 & 6.35 & 7.19 & 6.67 \\
 & Slowdown & 31 & 2.59 & 1.62 & 1.65 & 1.66 \\
\specialrule{0.3pt}{2pt}{2pt}
\multirow[c]{2}{*}{Extended} & Latency & 64 & 5.94 & 5.45 & 7.46 & 7.17 \\
 & Slowdown & 62 & 2.69 & 1.80 & 1.38 & 1.44 \\
\bottomrule
\end{tabular}%
}
\end{table}

\noindent\textbf{Analysis.}
Table~\ref{tab:contention-group-slowdown} compares isolated and concurrent latency for requests with 128 input and 128 output tokens. Group B's measured TTFT rises from 5.588 to 6.411\,s (14.7\%), while its TPOT changes by less than 1\,ms. Enabling contention raises B's predicted TTFT from 5.285 to 5.616\,s, reducing the prediction error from 1.125 to 0.795\,s. For group A, the same correction moves the prediction further from the measurement. These results show the importance of modeling contention and suggest that network contention is more pronounced during prefill than during decoding.

\begin{table}[t]
\centering
\caption{Latency for groups A, B, and C with 128 input and 128 output tokens. Predictions are for concurrent execution with cross-group contention disabled (Off) or enabled (On).}
\label{tab:contention-group-slowdown}
\footnotesize
\setlength{\tabcolsep}{3pt}
\resizebox{0.8\columnwidth}{!}{%
\begin{tabular}{@{}ll rr rr@{}}
\toprule
\multirow[c]{2}{*}{Metric} & \multirow[c]{2}{*}{Group}
 & \multicolumn{2}{c}{Measured} & \multicolumn{2}{c}{Predicted} \\
\cmidrule(lr){3-4}\cmidrule(lr){5-6}
 & & Isolated & Concurrent & Off & On \\
\midrule
\multirow[c]{3}{*}{TTFT (s)} & A & 10.842 & 10.943 & 11.723 & 12.040 \\
 & B & 5.588 & 6.411 & 5.285 & 5.616 \\
 & C & 10.535 & 10.561 & 10.737 & 10.737 \\
\specialrule{0.3pt}{2pt}{2pt}
\multirow[c]{3}{*}{TPOT (ms)} & A & 308.0 & 305.9 & 310.2 & 310.6 \\
 & B & 256.8 & 256.1 & 221.9 & 223.7 \\
 & C & 438.5 & 431.6 & 434.9 & 434.9 \\
\bottomrule
\end{tabular}%
}
\end{table}


\end{document}